\documentclass[sigplan, nonacm]{acmart}
\AtBeginDocument{%
  \providecommand\BibTeX{{%
    \normalfont B\kern-0.5em{\scshape i\kern-0.25em b}\kern-0.8em\TeX}}}

\setcopyright{acmlicensed}
\copyrightyear{2018}
\acmYear{2018}
\acmDOI{XXXXXXX.XXXXXXX}

\acmISBN{978-1-4503-XXXX-X/18/06}

\usepackage{multirow}
\usepackage{hyperref}
\hypersetup{
    colorlinks = true,
    linkcolor = blue,
    urlcolor = blue,
}

\usepackage{xcolor}
\definecolor{linkblue}{RGB}{0,112,193}
\newcommand{\link}[1]{\textcolor{linkblue}{#1}}

\definecolor{functioncolor}{RGB}{197,134,51}
\definecolor{paramcolor}{RGB}{0,112,193}
\definecolor{docstringcolor}{RGB}{35,142,35} 
\definecolor{commentcolor}{RGB}{206,145,120}

\usepackage{tcolorbox}
\newcommand{\grayboxblue}[2]{%
\begin{tcolorbox}[width=0.99\linewidth, colback=gray!05, colframe=cyan!50!blue, boxrule=1pt, arc=3pt, fonttitle=\small\bfseries, title={#2}]
\small\texttt{#1}
\end{tcolorbox}
}

\usepackage{xcolor} % Package for defining and using colors
\definecolor{jsonkeycolor}{RGB}{197,134,51} 
\definecolor{stringcolor}{RGB}{35,142,35}

\usepackage{pdfpages}

\begin{document}

\title{DocuMint: Docstring Generation for Python using Small Language Models}

\author{Bibek Poudel*, Adam Cook*, Sekou Traore*, Shelah Ameli*} 
\authornote{Authors contributed equally to this research.}

\affiliation{ 
  \institution{Min H. Kao Department of Electrical Engineering and Computer Science}
  \city{University of Tennessee, Knoxville} 
  \state{TN} 
  \country{USA}
}
\email{{bpoudel3,acook46,staore1,oameli}@vols.utk.edu}
    
% \affiliation{%

% \affiliation{%

% \affiliation{%

% \affiliation{%


\begin{abstract}
Effective communication, specifically through documentation, is the beating heart of collaboration among contributors in software development. Recent advancements in Language Models (LMs) have enabled the introduction of a new type of actor in that ecosystem: LM-powered assistants capable of code generation, optimization, and maintenance. Our study investigates the efficacy of Small Language Models (SLMs) in generating high-quality docstrings by assessing accuracy, conciseness, and clarity, benchmarking performance quantitatively through mathematical formulas and qualitatively through human evaluation using Likert scale. Further, we introduce DocuMint, as a large-scale supervised fine-tuning dataset with $100,000$ samples. In quantitative experiments, Llama 3 $8$B achieved the best performance across all metrics, with conciseness and clarity scores of $0.605$ and $64.88$, respectively. However, under human evaluation, CodeGemma $7$B achieved the highest overall score with an average of $8.3$ out of $10$ across all metrics. Fine-tuning the CodeGemma $2$B model using the DocuMint dataset led to significant improvements in performance across all metrics, with gains of up to $22.5\%$ in conciseness. The fine-tuned model and the dataset can be found in HuggingFace\footnote{\link{\textbf{\underline{\url{https://huggingface.co/documint}}}}} and the code can be found in the repository\footnote{\link{\textbf{\underline{\url{https://github.com/Docu-Mint/DocuMint}}}}}.

\end{abstract}





\maketitle

\section{Introduction}
\label{sec:intro}
% The latest saga in this oddessy is the genesis of
% Nevertheless, Code LLMs exhibit the potential to enhance all phases of the software development cycle~\cite{lozhkov2024starcoder}, such as Design2Code, which automates Front End Engineering~\cite{si2024design2code}. 
% Code is the frontier of Artificial Intelligence. This has lead to the emergence of Large Language Models for Code (Code LLMs) including foundation models such as CodeLlama~\cite{roziere2023code} and StarCoder~\cite{lozhkov2024starcoder} as well as products such as Copilot~\cite{github_copilot} and GhostWriter~\cite{replit_ai}. While a formal syntax, explicit structure, limited vocabulary, and deterministic outcomes in coding provide an ideal environment for LLMs to prove useful, there also exist challenges in understanding the context, natural language instruction, and code comprehension. Nevertheless, Code LLMs exhibit the potential to enhance all phases of the software development cycle~\cite{lozhkov2024starcoder}. Such as Design2Code, automate Front End Engineering~\cite{si2024design2code}. we have entered into an era of LLM enabled AI agents which can automate the entire software engineering pipeline. As the LLM landscape becomes more agent-based with software engineer agents such as Devin~\cite{cognition_labs}, Devika~\cite{devika_github}, and OpenDevin~\cite{opendevin_github}; software development is done with a mixture of human and AI input, it becomes even more important for humans and AI to understand each other's intent and comprehend the code written by each other.
% generate comprehensible code
Code is at the frontier of Artificial Intelligence (AI). The emergence of Large Language Models for Code (Code LLMs), including foundation models such as CodeLlama~\cite{roziere2023code}, Code-\newline Gemma~\cite{code_gemma}, and StarCoder~\cite{lozhkov2024starcoder}, along with products like GitHub Copilot~\cite{github_copilot} and GhostWriter~\cite{replit_ai}, has paved the way for even more advanced software development tools. The latest saga in this odyssey is the rise of CodeLLM-enabled AI agents such as Devin~\cite{cognition_labs}, Devika~\cite{devika_github}, and OpenDevin~\cite{opendevin_github}, which can automate entire software development pipelines, ushering us into an era where software is co-developed by humans and AI. Already in $2022$, GitHub reported that on average, $46\%$ of all code written across all programming languages was assisted by GitHub Copilot~\cite{github_copilot_business}. While the formal syntax, limited vocabulary, and deterministic outcomes in code provide an ideal (structured) environment for CodeLLMs to thrive~\cite{zhang2023unifying}, at the same time, their usefulness is challenged by the need to understand context, interpret natural language instructions, and convey the intent for generated code. As the CodeLLM landscape evolves and software development increasingly involves collaboration between human and AI, it becomes increasingly important for both parties to convey their objective and comprehend the code generated by their counterparts.

\begin{figure}[t!]
\vspace{10pt}
\centering
\scalebox{0.90}{
    \begin{minipage}{1.0\linewidth}
        \begin{tcolorbox}[colback=white, colframe=white, boxrule=0.5pt, arc=3pt]
            \small{\ttfamily{\fontseries{b}{\selectfont{
            \textcolor{paramcolor}{def} \textcolor{functioncolor}{example\_function}(\textcolor{paramcolor}{param1}\textcolor{black}{,} \textcolor{paramcolor}{param2\vspace{6pt}}\textcolor{black}{):}\\
            \hspace*{8mm}\textcolor{docstringcolor}{"""}\\
            \hspace*{8mm}\textcolor{docstringcolor}{This is an example of a docstring}\\
            \hspace*{8mm}\textcolor{docstringcolor}{for a Python function.}\\
            \hspace*{8mm}\textcolor{docstringcolor}{}\\
            \hspace*{8mm}\textcolor{docstringcolor}{Docstrings are enclosed in triple}\\
            \hspace*{8mm}\textcolor{docstringcolor}{quotes and appear immediately}\\
            \hspace*{8mm}\textcolor{docstringcolor}{after the function definition.\vspace{4pt}}\\
            \hspace*{8mm}\textcolor{docstringcolor}{"""\vspace{-4pt}}\\
            \hspace*{8mm}\\
            \hspace*{8mm}\textcolor{commentcolor}{\#~Function implementation}
            }}}}
        \end{tcolorbox}
    \end{minipage}
}
\vspace{-10pt}
\caption{\small{An example of docstring for a python function. Docstrings serve as documentation and help developers understand how to use the respective code.}}
\label{fig:docstring_example}
\vspace{-28pt}
\end{figure}

Given the impressive capabilities of LLMs and CodeLLMs, there is growing interest in smaller, more efficient models called Small Language Models (SLMs)~\cite{microsoft_phi_2}, such as Phi~\cite{abdin2024phi3}, Gemma~\cite{google2024gemma} and smaller variants of Llama~\cite{meta_llama_2023, llama3modelcard}. SLMs are significantly more cost-effective in terms of training and inference (latency, memory, throughput, and energy consumption), and they are small enough for considerations on running locally; about $7$B parameters for consumer-level GPUs and about $2$B parameters for comsumer CPUs. \textbf{As language models continue to shape the landscape of software development, high-quality documentation becomes increasingly crucial for facilitating effective communication and collaboration between humans and AI}. Given that developers already spend $58\%$ of their time on comprehension of unfamiliar code~\cite{xia2017measuring}, writing clear and consistent documentation helps helps reduce this time while enabling both human developers and AI to convey the intent behind the code and understand each other's contributions better. This not only enhances the developer experience but also improves overall software quality.

Although CodeLLMs are capable of generating docstrings (as docstrings are a part of their pre-training and fine-tuning data), to date, limited research has focused on evaluating the quality of the generated docstrings. Existing benchmarks, such as HumanEval~\cite{chen2021evaluating} and MBPP~\cite{austin2021program}, assess code generation performance, leaving a gap in measuring the quality of the accompanying documentation. Furthermore, traditional natural language generation metrics, such as Bilingual Evaluation Understudy (BLEU), fall short in evaluating docstring quality due to their emphasis on n-gram overlap between generated and reference sentences, disregarding the semantic meaning and context. In this work, we address these limitations by introducing DocuMint, a large-scale dataset and methodology for evaluating the quality of docstrings generated by SLMs. We make three key contributions:
\vspace{0pt}
\begin{itemize}
    \item First, we breakdown the ``quality" assessment of docstrings into three categories: accuracy, conciseness, and clarity with well defined metric for each category. 
    \item Second, we benchmark the performance of leading code generation SLMs in generating docstrings using both the mathematical metrics and human evaluation. 
    \item Third, we introduce DocuMint, a supervised fine-tuning dataset consisting of $100,000$ samples.
\end{itemize}

To the best of our knowledge, DocuMint is the first comprehensive work designed to evaluate and improve the quality of docstrings generated by SLMs, providing a large-scale dataset, benchmarks, a fine-tuned model, and an evaluation methodology.

% More specifically what is the task?
% - Beginner friendly documentation
% - Small Language Model i.e., Local GPU implementation

% \begin{figure}[t!]
% \vspace{12pt}
% \centering
% \grayboxblue{%
% \textmd{def example\_function(param1, param2):\newline\hspace*{6mm}"""\newline\hspace*{6mm}This is an example of a docstring\newline\hspace*{6mm}for a Python function.\newline\hspace*{6mm}\newline\hspace*{6mm}Docstrings are enclosed in triple\newline\hspace*{6mm}quotes and appear immediately\newline\hspace*{6mm}after the function definition.\newline\hspace*{6mm}"""\newline\hspace*{6mm}\newline\hspace*{6mm}\#~Function implementation}}{Docstring Example}
% \vspace{-6pt}
% \caption{An example of a Python function with a docstring. Docstrings serve as documentation for the function and help users understand how to use it effectively.}
% \label{fig:docstring_example}
% \end{figure}

% 1) No such large-scale, well-put, conference, ... (eval) 

% 2) BLEU (Metrics) -- IEEE Standard / Accuracy

% 3) Fine-tune to improve performance from levereging dataset (train dataset)

%%%%%%%%%%%%%%%%%%%
% Large Language Models (LLMs) such as GPT-4~\cite{openai_gpt4_2023} and Gemini~\cite{google_gemini_2024} represent a significant advancement in artificial intelligence.

% SLMs have a limited capacity for knowledge representation and reasoning however, despite their size, SLMs can be useful in certain scenarios where the specific knowledge encoded in the model is relevant to the task at hand. Making them ideal for 
\section{Related Work}
\label{sec:related}
% While these works are important, they provide little utility to software engineering hence will be excluded.
There exist two mainstream code-related evaluations for LLMs: code generation and code comprehension.

\textbf{Code Generation:} The standard code-related evaluation benchmarks for language models, such as HumanEval~\cite{chen2021evaluating}, MBPP~\cite{austin2021program}, and SWE-Bench~\cite{jimenez2023swe}, assess their abilities to solve problems through generated code. HumanEval presents carefully crafted programming problems and evaluates whether the generated code solutions pass on hidden test cases. MBPP tests the ability of LLMs to translate instructions into functional code. More recently, SWE-Bench tests their abilities to tackle real-world software issues sourced from GitHub, directly assessing a model's proficiency in understanding and resolving software problems.

\textbf{Code Comprehension:} Most of the prior research in this field has focused on using LLMs to comprehend human-written or AI-generated code, including code summarization~\cite{cai2024fly, wang2024demo2code} and explanations~\cite{sarsa2022automatic}. These tasks leverage the abilities of LLMs to reason in a Chain-of-thought~\cite{wei2022chain} or in a step-by-step~\cite{kojima2022large} manner. Code summarization focuses on generating succinct, human-readable summaries of code snippets, whereas step-by-step explanations are more prominent in educational settings~\cite{brusilovsky2023explaining}. 

To date, limited research has focused on the evaluation of CodeLLMs in docstring generation, and prominent datasets for this task are scarce. Early works on code documentation generation, such as~\cite{barone2017parallel}, introduced datasets of parallel code and natural language descriptions of $150,370$ Python function declarations, bodies, and docstrings scraped from open-source repositories. However, these efforts were limited by the size and diversity of the available data, and they were conducted before the emergence of language models, which questions their relevance. Similarly, line-by-line comment generation~\cite{hu2018deep} and contextual function/method call and usage information~\cite{mcburney2014automatic} have been explored, but these works also predate the rise of LLMs. More recently, RepoAgent~\cite{luo2024repoagent} has demonstrated the potential of LLMs for repository-level code documentation generation, generating and proactively maintaining high-quality documentation for entire projects.

While these existing works provide valuable insights, they differ from our approach, which is to evaluate the abilities of CodeLLMs themselves to generate documentation alongside the code they produce. We evaluate the efficacy of language models in documentation generation as a software development tool, i.e., \textbf{given a code function, is the CodeLLM able to describe its core functionality, input/output parameters, and intended usage in an accurate, concise, and clear manner?}

%Although code comprehension tasks may appear similar to our work because they also involve code input and natural language output
% limited research has focused on evaluation of CodeLLMs in docstring generation, neither do prominent such datasets exist. Early works on code documentation generation, such as~\cite{barone2017parallel}, introduced datasets of parallel code and natural language descriptions of $150,370$ Python function declarations, bodies, and docstrings scraped from open-source repositories on GitHub. However, these efforts were limited by the size and diversity of the available data also, as this happened before the emergence of LLMs. For training they used neural machine translation models and used BLEU for metric. However, our focus in .. Can CodeLLMs explain the code they write? leveraging SLMs, newer metrics.  

% Iterative code refinement~\cite{zheng2024opencodeinterpreter}

%IEEE Standard.
\section{Preliminaries}
In this section, we introduce the various training phases of a language model and the tasks that it is trained to perform. 

\label{sec:prelim}
\subsection{Training}
Creating a helpful Language Model (LM) involves three essential steps: pre-training, fine-tuning, and alignment.

\textbf{Pre-training:} Pre-training involves training a LM on vast amounts of unlabeled text data using self-supervised learning techniques such as masked language modeling or next token prediction. This process allows the model to capture the intrinsic patterns and statistical regularities present in natural language. Pre-training is computationally intensive and requires clusters of GPUs to process the internet-scale data efficiently, resulting in a "base model" that serves as a foundation for further fine-tuning. 

\textbf{Fine-tuning:} Pre-training alone does not enable the model to understand or follow specific instructions. The utility of LMs is significantly improved by fine-tuning and depending on the nature of the task, fine tuning could be Supervised Fine Tuning (SFT) or Instruction Tuning (IT). SFT adapts a pre-trained base model to perform specific downstream tasks by training it on labeled data relevant to the task. Whereas IT is employed to turn the base model into a useful assistant capable of understanding and responding to user instructions expressed in natural language.

%i.e., the model is exposed to a diverse range of instructions and their corresponding desired outputs which helps the model learn to interpret the semantics of the instructions and generate responses that align with the instructions.

\textbf{Alignment:} Also known as human preference fine-tuning, this step aligns the behavior of an LM with human preferences, enhancing its friendliness, helpfulness, and safety. This process involves collecting human feedback on the model's outputs and using this feedback as a reward signal to guide the model's behavior. By optimizing the model to generate outputs that align with human preferences, this fine-tuning helps to mitigate undesirable behaviors and ensures more socially appropriate and beneficial responses.

\subsection{Tasks}
\noindent LMs can be useful for various tasks, such as text/code infilling and completion.

\textbf{Infilling:} is a specialized task designed for code-generation models, where the objective is to generate code snippets or comments that best fit within a given prefix and suffix. This task is particularly relevant for code assistants, which are trained to provide code suggestions based on the surrounding context at the current cursor position. Infilling models learn to capture the syntactic and semantic patterns present in code, allowing them to generate coherent and contextually appropriate code fragments.

\textbf{Completion:} refers to the fundamental capability of LMs to generate fluent and coherent text by predicting the next token in a sequence based on the preceding context. Completion allows LLMs to generate human-like text that follows the style, tone, and subject matter of the given prompt. This is the most common task of a LM.

%This capability is acquired during the pre-training phase, where the model learns the statistical regularities and patterns present in natural language.

% \noindent Benchmarking LLMs ..

% \textbf{BLEU:} calculating the similarity between generated sentences and reference sentences.. unable to decide the functional equivalence between generated code and reference code (but we are looking at docstring) 
\section{Methodology}
\label{sec:method}
In this section, we provide an overview of the selected state-of-the-art models, the data used for inference and bench-marking them, the data extraction process for fine-tuning, and the applied fine-tuning technique.

\subsection{Selected Models}

Models were selected from the EvalPlus Leaderboard~\cite{liu2023your} at full precision, with their rank on the leaderboard as of April $25$,$2024$ shown in Table\ref{table:models}.

% \textbf{CodeGemma~$2$B}~\cite{code_gemma} - \#83

% \textbf{CodeGemma~$7$B Instruct}~\cite{code_gemma} - \#43

% \textbf{DeepSeek Coder~$1.3$B} - \#74

% \textbf{DeepSeek Coder~$6.7$B Instruct} - \#14

% \textbf{StarCoder2~$2$B} - \#71

% \textbf{StarCoder2~$7$B} - \#66

% \textbf{Meta Llama3~$8$B Instruct} - \#38

\vspace{-10pt}
\begin{table}[h!]
\begin{center}
\vspace{0pt}
  \normalsize
  \setlength{\tabcolsep}{12pt}
  \scalebox{1.0}{
  \begin{tabular}{lc}
    \toprule
    Model & Rank \\
    \hline   
    DeepSeek Coder~$6.7$B Instruct & $14$ \\
    Meta Llama3~$8$B Instruct & $38$\\
    CodeGemma~$7$B Instruct & $43$ \\
    StarCoder2~$7$B & $66$ \\
    %StarCoder2~$2$B & $71$ \\
    %DeepSeek Coder~$1.3$B & $74$\\
    CodeGemma~$2$B & $83$ \\
    \bottomrule
  \end{tabular}}
\end{center}
\vspace{0pt}
\caption{\small{Selected models and their rank.}}
\vspace{-18pt}
\label{table:models}
\end{table}

Since the selected models are CodeLLMs, use instruct variants whenever available.

\subsection{Benchmarking Data}

For benchmarking the selected models, we extract seven python functions from three widely established NLP datasets~\cite{chang2023survey, zhao2023survey, roziere2023code, nijkamp2022codegen, li2022competition, wei2022chain, kaddour2023challenges} and perform model inference.

The Mostly Basic Python Problems \textbf{MBPP} dataset~\cite{austin2021program} consists of around $1,000$ crowd-sourced Python programming problems, designed to be solvable by entry level programmers, covering programming fundamentals, standard library functionality, and so on. Each problem consists of a task description, code solution and $3$ automated test cases. The dataset is split into a train and test set, we select functions from the test set.

The Hand-Written Evaluation Set \textbf{HumanEval} dataset~\cite{chen2021evaluating} consists of $164$ hand written problems with a function signature, a docstring describing the task, reference solution, and tests. These problems are our medium level functions, they contain functions and problems that are notably more complex than the MBPP dataset but less algorithmic intensive when compared to our selected APPS functions.

The Automated Programming Progress Standard \textbf{APPS} dataset~\cite{hendrycks2021measuring} comprises 10,000 coding examples, evenly distributed between train and test sets. These examples were sourced from various open-access coding websites like Codeforces and Kattis. Each problem within the dataset includes a question description, solutions, input/output details, difficulty level, and source URL, with problems categorized into Introductory (simple and easy), Interview (technical interview questions), and Competition Level (advanced programming competitions).

We focused on functions categorized under the 'interview level' section of the test set. From each problem, we extracted solutions contained within functions defined by the 'def' keyword, excluding class objects from our inference dataset for analysis.

\subsection{Fine Tuning Data}

As the human annotation process is expensive and relatively slow, we leverage existing code from the Free and open-source software (FOSS) ecosystem. We utilize World of Code (WoC) ~\cite{ma2019world, ma2021world}, an efficient and flexible project analysis framework that provides an abstracted interface to the intricacies of this ecosystem, enabling a reliable and straightforward approach to research. Our goal is to ensure quantity, quality, and diversity of data for fine-tuning.

We query the repositories in WoC and filter them based on metrics such as number of contributors $(>50)$, commits $(>5k)$, stars $(>35k)$, and forks $(>10k)$. This ensures that we focus on well-established and actively maintained projects. To extract functions from the filtered repositories, we employ a parser that utilizes the GitPython and abstract syntax tree modules to navigate through repository files and identify function definitions. The parser extracts $100,000$ Python functions along with their respective docstrings. The data was formatted into a JSON file with the Alpaca instruction format~\cite{taori2023alpaca} shown below.

\begin{figure}[h!]
\vspace{-10pt}
\centering
\scalebox{0.90}{
    \begin{minipage}{1.0\linewidth}
        \begin{tcolorbox}[colback=white, colframe=white, boxrule=0.5pt, arc=3pt]
            \small{\ttfamily{\fontseries{b}{\selectfont{
            \textcolor{jsonkeycolor}{\{}\\
            \textcolor{jsonkeycolor}{"instruction"}: \textcolor{stringcolor}{"<function definition>"},\vspace{6pt}\\
            \textcolor{jsonkeycolor}{"response"}: \textcolor{stringcolor}{"<ideal docstring>"}\\
            \textcolor{jsonkeycolor}{\} }\\
            }}}}
        \end{tcolorbox}
    \end{minipage}
}
\vspace{-26pt}
\end{figure}

% In addition techniques such as QLoRA~\cite{dettmers2024qlora} builds upon LoRA by further reducing the memory footprint through quantization. (does not sacrifice performance compared to standard 16-bit finetuning or LoRA finetuning methods.)
% precision ($32$ bit) training

\subsection{Fine Tuning}

The full model fine-tuning for language models, even for modest sizes, is memory and compute-intensive; hence, we resort to parameter-efficient fine-tuning, updating only a small subset of model parameters, among which Low-Rank Adaptation (LoRA)\cite{hu2021lora} is a popular technique. We perform supervised instruction fine-tuning on the Gemma-2B base model, which has been trained to predict the next token on internet text without any instructions, using LoRA. This fine-tuning process involves using examples that demonstrate how the model should respond to specific instructions. The total fine-tuning time was $48$ GPU hours, with $78,446,592$ LoRA parameters at full precision ($32$ bit) and $185,040,896$ total training tokens. The full set of parameters used in fine-tuning are provided in Table\ref{table:params}.

% \graybox{%
% \small{\texttt{You are a helpful AI assistant that specializes in generating high-quality docstrings for Python code functions. Your task is to create docstrings that are:\newline\newline Concise: Brief and to the point, focusing on essential information.\newline\newline Complete: Cover functionality, parameters, return values, and exceptions.\newline\newline Clear: Use simple language and avoid ambiguity.}}}%

\begin{table}[t!]
\begin{center}
\vspace{0pt}
  \normalsize
  \setlength{\tabcolsep}{5pt}
  \scalebox{1.0}{
  \begin{tabular}{llc}
    \toprule
    Category & Parameter& Value \\
    \hline    
    \multirow{3}{*}{CodeGemma} 
    & vocabulary size & $256000$ \\
    & max pos. embeddings & $8192$ \\
    Config& num\_hidden\_layers & $18$ \\
    & num\_heads & $8$ \\
    \hline    
    \multirow{7}{*}{Fine} 
    & max\_seq\_len & $128$ \\
    & LoRA rank & $64$ \\
    & LoRA alpha& $128$ \\
    & LoRA dropout& $0.1$ \\
    Tuning& batch size & $8$ \\
    & grad accumulation steps &$16$\\
    & epochs & $4$ \\
    % & num\_activations & - \\
    & optimizer & $8$ bit adam\\
    & Learning rate (initial)& $2$e$-4$\\

    \bottomrule
  \end{tabular}}
\end{center}
\vspace{2pt}
\caption{\small{Detailed configuration of the CodeGemma model and hyper-parameters used in fine-tuning.}}
\vspace{-24pt}
\label{table:params}
\end{table}

\section{Experiments}
\label{sec:exp}

In this section, we define our experiments, describe the metrics used to assess the generated docstrings, provide summary statistics of the human participants, and detail the experiment setup.\\ 

\noindent Our research objectives are contained in the three experiments described below.

\textbf{Experiment $1$} quantitatively evaluates the docstrings generated by SLMs using the three metrics: Accuracy, Conciseness, and Clarity, as described in Section~\ref{section:metrics}.

\textbf{Experiment $2$} qualitatively evaluates a subset of the docstrings generated by SLMs using a human questionnaire where participants with prior Python programming experience score the Accuracy, Conciseness, and Clarity of generated docstrings on a Likert scale.

\textbf{Experiment $3$} evaluates the docstrings generated by a SLM that has been fine-tuned on data extracted from World of Code and quantitatively compares (similar to Experiment $1$) the performance against the original model.

%i.e., quantitatively ranks the SLMs across the metric category (with a control LLM used as reference in Accuracy).
%evaluates the metrics as a Likert scale, along with an additional metric for overall quality of the docstring.
%using the same techniques as detailed in Experiment 1.

% \begin{table*}[h!]
% \vspace{8pt}
% \centering
% \setlength{\tabcolsep}{15pt} 
% \renewcommand{\arraystretch}{1.5} 
% \begin{center}
% \scalebox{1.0}{
%     \begin{tabular}{|l|cc|cc|cc|}
%     \hline
%         \multirow{2}{*}{Model} & \multicolumn{2}{c|}{Completeness} & \multicolumn{2}{c|}{Conciseness} & \multicolumn{2}{c|}{Clarity}\\ 
%         \cline{2-7} 
%          & Human & Math & Human & Math & Human & Math\\ 
%         \hline
%         Code Gemma $7$B & $-$ & $-$ & $-$ & $-$ & $-$ & $-$\\
%         % \cline{1-6}
%         Code Llama $7$B& $-$ & $-$ & $-$ & $-$ & $-$ & $-$\\
%         Deepseek Coder $6.7$B & $-$ & $-$ & $-$ & $-$ & $-$ & $-$\\
%         StarCoder-v$2$ $7$B  & $-$ & $-$ & $-$ & $-$ & $-$ & $-$\\
%         \hline
%     \end{tabular}}
% \end{center}
% \vspace{-6pt}
% \caption{ [Collect the results in the table but for final publishing, Replace this with a plot] Comparative Evaluations of various CodeLLMs in generating docstring as measured by .. bfloat16 data-type with a batch size of 1}
% \end{table*}

\begin{table*}[h!]
\vspace{8pt}
\centering
\setlength{\tabcolsep}{15pt} 
\renewcommand{\arraystretch}{1.3} 
\begin{center}
\scalebox{1.0}{
    \begin{tabular}{lccc}
        \toprule
        Model & Accuracy & Conciseness & Clarity\\ 
        \hline
        CodeGemma 7B & $0.609$ & $0.569$ & $76.49$ \\
        DeepSeek Coder $6.7$B & $\bf{0.679}$ & $0.734$ & $\bf{64.44}$ \\
        Llama3 8B & $0.668$ & $\bf{0.605}$ & $\bf{64.88}$  \\
        StarCoder$2$ $7$B& $0.626$ & $0.510$ & $69.74$ \\
        \bottomrule
    \end{tabular}}
\end{center}
\vspace{4pt}
\caption{Experiment 1 results i.e., comparative evaluation of various CodeLLMs in generating docstring as measured by Accuracy, Conciseness, and Clarity. Instruct variants used for CodeGemma, DeepSeek, and Llama3. Overall, Llama 3 performs the best across all three metrics. DeepSeek achieves the best Accuracy score, CodeGemma achieves the best Conciseness score, and both DeepSeek and Llama 3 share the best Clarity score.}
\vspace{-16pt}
\label{exp:first}
\end{table*}

\begin{figure}[t!]
\centering
\grayboxblue{%
You are a helpful AI assistant that specializes in generating high-quality docstrings for Python code functions. Your task is to create docstrings that are: \newline\newline Accurate: Cover functionality, parameters, return values, and exceptions. \newline\newline Concise: Brief and to the point, focusing on essential information.\newline\newline Clear: Use simple language and avoid ambiguity.\newline\newline Generate docstring in this format: """<generated docstring>""".}{System Prompt}
\vspace{-10pt}
\caption{System prompt that is attached before the function definition when querying the model for inference.}
\vspace{-14pt}
\label{fig:system_prompt}
\end{figure}

\subsection{Experiment Metrics}
\label{section:metrics}
Classic NLP metrics such as Bilingual Evaluation Understudy (BLEU) are often used to evaluate machine translation systems by comparing the machine-generated translations to human-generated reference translations. However, they are not reliable in the context of code~\cite{evtikhiev2023out}. Two dramatically different code snippets can be semantically equivalent, and since the same vocabulary is often used with docstrings attached to the code, we refrain from using BLEU. We consider the primary concern when it comes to LLM-generated docstring is accuracy (correctness); to further get a comprehensive measurement, we consider conciseness, and clarity.

\textbf{Accuracy:} measures the correctness in the description of the function logic and coverage of the generated docstring on associated code elements such as input and output variables. We measure accuracy using BERTScore~\cite{zhang2019bertscore} given by

$$
    \text{Accuracy} = \frac{v_g \cdot v_e}{\sqrt{v_g^2} \cdot \sqrt{v_e^2}},
$$

\noindent where $v_g$ and $v_e$ are the Bidirectional Encoder Representations from Transformers~\cite{devlin2018bert} encodings of the SLM generated docstring and an ``expert'' docstring generate by the Claude-3 Opus~\cite{anthropic2024claude3} LLM, respectively. A higher accuracy score indicates a greater similarity  between the generated docstring and expert docstring and vice versa. 

\textbf{Conciseness:} measures the ability of the generated docstring to convey information succinctly without unnecessary verbosity. We measure conciseness using a compression ratio defined as:
$$
\text{Conciseness} = \frac{\text{Text}_C}{\text{Text}_O},
$$

\noindent where ${\text{Text}_C}$ and ${\text{Text}_O}$ are the sizes of the compressed text and the original text respectively. The compression ratio is bounded between $0$ and $1$ with a lower ratio indicating more conciseness. An ideal score is between $0.5$ and $0.6$ i.e., if the score is too low, it indicates that the generated docstrings are extremely short.

\textbf{Clarity:} measures the readability of the generated docstring i.e., uses simple language and is unambiguous. We measure clarity using the Flesch-Kincaid reading score:

$$
\text{Clarity} = 206.835 - (1.015 * \frac{w}{l}) - (84.6 * \frac{s}{w})),
$$    

\noindent where w is the number of words generated, l is the number of sentences or lines, and s is the number of syllables. A score between $70-100$ indicates that the text is very easy to read, at $50-70$ text slowly becomes more difficult to read (high school level), and $0-50$ is regarded as difficult to read and best understood by university graduates. Ideally, for docstrings a score of $50-70$ is desired.

\subsection{Summary statistics of the Human Population}

We selected participants to perform a `qualitative' measure of the docstrings generated by each of the four Small Language Models (SLMs). Each participant evaluated five questions from the inference dataset, with three persons per SLM. A sample of the questionnaire is included in appendix \ref{appendix} for reference.

\begin{figure}[h!]
    \centering
    \includegraphics[width=3in]{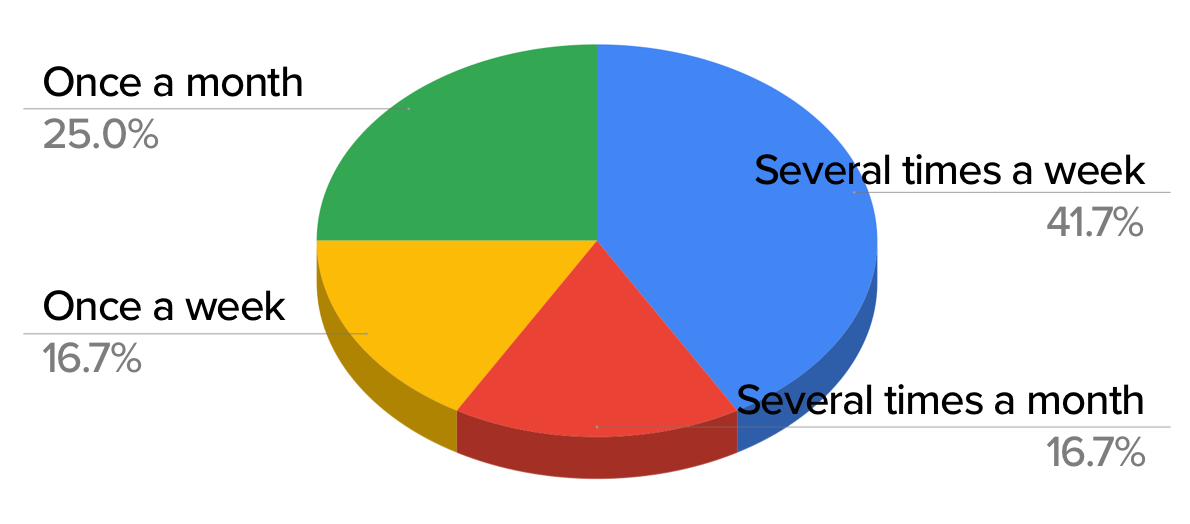}
    \vspace{-10pt}
    \caption{Survey participants' frequency in reading docstrings. About 42$\%$ read docstrings several times a week,  $\sim$  17$\%$ read docstrings once a week and several times a month each, while a quarter of them reads docstring once a month.}\label{exp_stats}
    \vspace{-10pt}
\end{figure}

The selected population of this study comprised of human programmers, with a median age of $25.5$ years. The cumulative years of Python Programming experience is 47 years, averaging 4 years per participant. We had a diverse group of 8 males and 4 females. Figure \ref{exp_stats} highlights the habits of the population with reading and engaging with docstrings.

%%%%%%%%%%%%

\subsection{Experiment Setup}

The data and model exploration phase was conducted using Lightning AI with A-$10$G GPUs. For inference by zero-shot prompting in experiment $1$, we employed an Intel $14900$K CPU, an Nvidia RTX-A$5000$ GPU, and $32$GB RAM. Fine-tuning experiments were performed using an Intel $12900$K CPU, an Nvidia RTX-$3090$ GPU, and $64$GB RAM. All models used in this study were obtained through HuggingFace.

% \subsection{Metrics}
% \textbf{Real World Evaluation}

% \subsection{Random content not organized yet}
% Experiment list: 

% 1. Main Experiment: Table below

% 2. Secondary experiment: Fine-tune vs no fine-tune using WoC data

% 3. Secondary experiment 2: If time permits: Emergent behavior study. (16 bit quant on 3, 7, 13 B models)

% For metrics: 
% 1. Something that we can measure.

% These things measure: ``helpfulness'', ``utility'', ``Beginner Friendliness''

% Experiment idea: No dataset exists for generating docstring for Python code. Run inference on the 100k fine tuning dataset. Find the hardest 100 when run inference on our models. Release the 100 as a dataset. 

% Emergent properties: One of the most interesting properties exhibited when models are scaled is that larger models tend to have abilities that smaller models do not, these abilities are often unpredictable. In this case, can they comprehend more complex functions..~\cite{wei2022emergent}

% Maybe we should use the PHI-2 model: 
% %https://www.microsoft.com/en-us/research/blog/phi-2-the-surprising-power-of-small-language-models/

% Phi-2 is also suitable for code and is already excellent in Coding benchmarks (surpasses larger models even a Llama-2 70B model)

\section{Experiment Results}
In this section, we present the results of our quantitative benchmarks, qualitative human questionnaires, and fine-tuning experiments on the selected SLMs.
\subsection{Experiment 1 Results}
%The values depicted in Table 3 are the numerical benchmarks for each SLM in the respective category.

Table~\ref{exp:first} shows the quantitative measurements of accuracy, conciseness, and clarity for the docstrings generated by SLMs, all of which have roughly 7 billion parameters. Accuracy scores were led by DeepSeek at $0.679$, with Llama 3 in second place at $0.668$. This indicates that the docstrings generated by DeepSeek and Llama 3 were the most similar to the control model (Claude 3 Opus, whose parameter count is in the hundreds of billions). On the other hand, CodeGemma scored the lowest accuracy at $0.609$, indicating the most dissimilarity with the control model.

For conciseness, CodeGemma had the best score at $0.569$, while StarCoder2, with a score of $0.510$, generated extremely short docstrings that may not convey enough information. Surprisingly, DeepSeek Coder exceeded the ideal threshold of $0.5-0.6$ with a score of $0.734$, implying that the generated docstrings were verbose. Further, DeepSeek, Llama3, and StarCoder2 have clarity scores within the desired range of $50-70$, with StarCoder2 at the upper threshold at $69.74$. However, CodeGemma exceeds the ideal threshold with a score of $76.49$. This means that while CodeGemma is deemed "easy to read," it uses much simpler language and syntax and may not be considered of professional quality.

\begin{figure}[t!]
    \centering
    \includegraphics[width=3in]{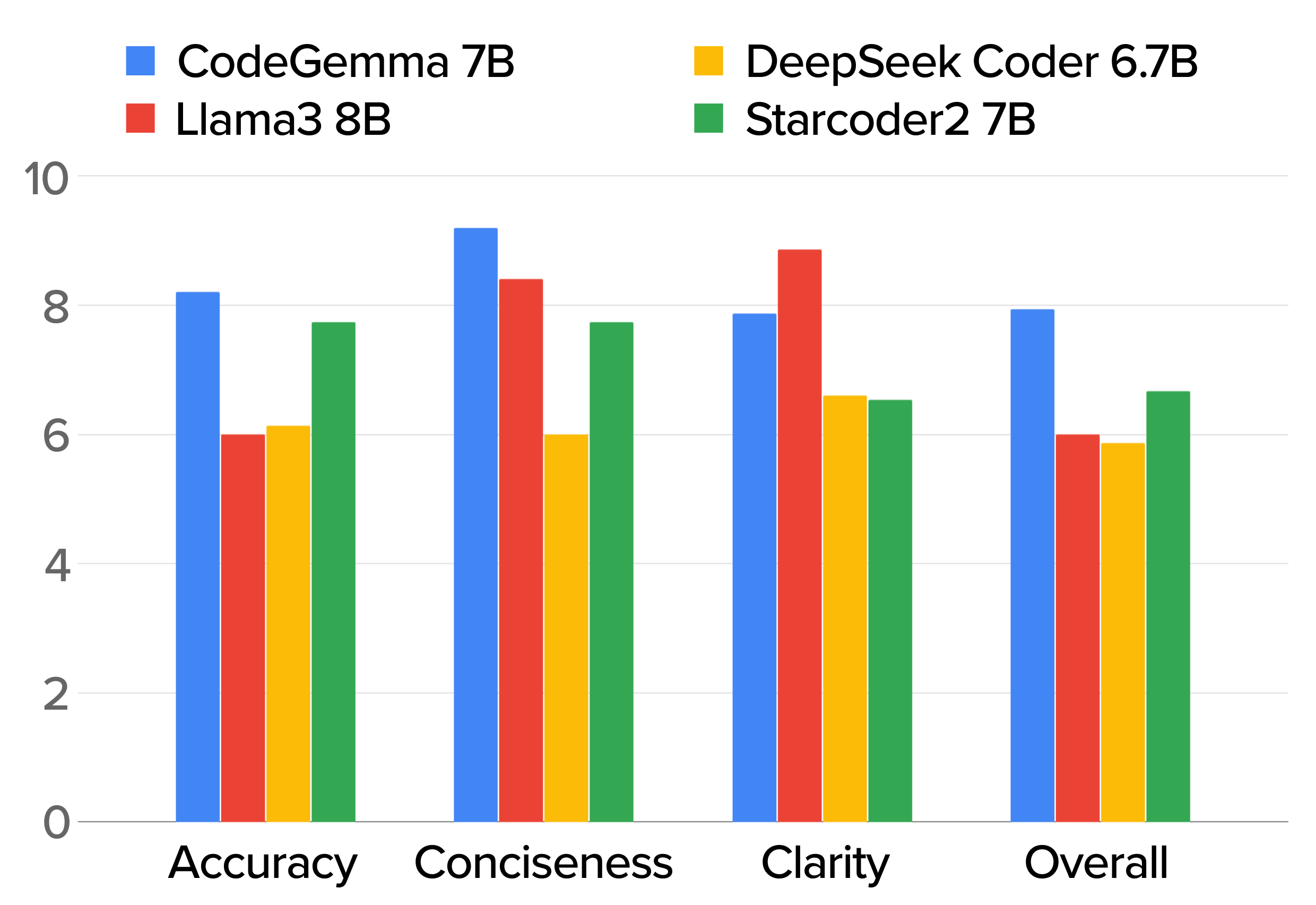}\\
    \caption{The average weighted scores from the human questionnaire. Instruct variants used for CodeGemma, DeepSeek, and Llama3. CodeGemma performs the best overall surpassing all other models in terms of accuracy and concisenes. Llama3 achieves the best clarity score. StarCoder2 was the second-best-performing model from a human perspective.}
    \label{exp2_results}
    \vspace{-10pt}
\end{figure}

\subsection{Experiment 2 Results}
Figure \ref{exp2_results} shows the average accuracy, conciseness, clarity, and overall scores from the human questionnaires. For accuracy, CodeGemma achieves the overall best performance with 8.2 out of 10. Surprisingly, Starcoder2 performed well with an accuracy of 7.7 from a human perspective despite having the worst results in the math benchmarks (Section~\ref{exp:first}). Llama3 and DeepSeek Coder have the lowest accuracy scores, which varies from the results achieved with the quantitative benchmarks. This suggests that humans consider CodeGemma and DeepSeek Coder as more reliable in generating correct docstrings compared to the other two models.

For conciseness, CodeGemma takes the lead with a high score of 9.2 out of 10, indicating its ability to generate more succinct and compact responses, this is consistent with the quantitative metric where CodeGemma also had the best conciseness score. Llama3 and Starcoder2 show similar levels of conciseness, while DeepSeek Coder generates the least concise score (5.8) among the four SLMs. Whereas for Clarity, humans rank Llama3 as being the clearest with a clarity score of $\sim$ 9, with CodeGemma coming closely behind with 7.8 clarity score demonstrating comparable performance. DeepSeek Coder and Starcoder2 exhibit lower clarity scores, suggesting that humans perceived their generated responses to be less understandable compared to the top two models.

Overall, CodeGemma and Starcoder2 stand out as the best-performing models, with CodeGemma having an overall score of 7.9 on the Likert scale. These two models consistently score higher across the three metrics compared to Llama3 and DeepSeek Coder. The results indicate that CodeGemma and Starcoder2 are more effective in generating accurate, concise, and clear responses.

% Each colored bar represents one the SLMs chosen: CodeGemma (blue), Meta Llama 3 (red), DeepSeek-Coder (yellow), and StarCoder2 (green)

\subsection{Experiment 3 Results}

\begin{figure}[h!]
        \centering
        \includegraphics[width=3in]{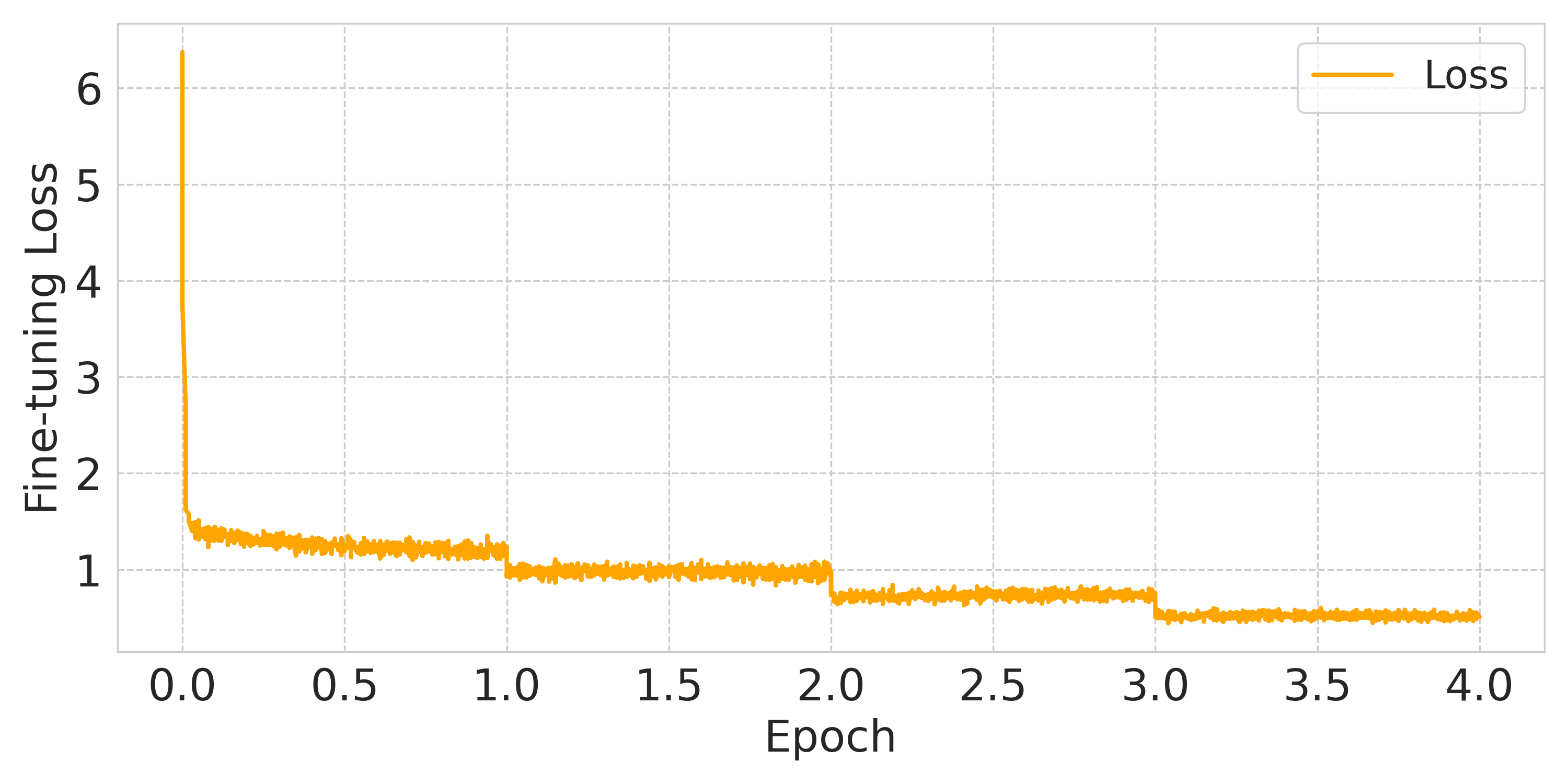}\\
        \vspace{-6pt}
        \caption{The loss curve during fine-tuning of the CodeGemma-$2$B model. The loss decreases sharply during the first epoch after which the curve saturates with marginal decrements per epoch.}\label{figure label}
        \vspace{-1pt}
        \label{figure:loss}
\end{figure}

Table~\ref{exp:third} presents the comparative evaluations between the base CodeGemma 2B model and its fine-tuned version. The fine-tuned model shows notable improvements across all metrics: accuracy increased by $12.7\%$, conciseness by $22.5\%$, and clarity by three reading levels. Despite the fine-tuned model being a 2B parameter model, it achieves competitive performance compared to the larger Llama3 $8$B model shown in Table~\ref{exp:first}. The fine-tuned CodeGemma 2B model has $14\%$ lower accuracy than Llama3 $8$B, but its conciseness falls within a desirable range, and its clarity is only slightly below the threshold. These results validate the effectiveness of our fine-tuning process using the data obtained from World of Code. The fine-tuning loss curve is shown in Figure~\ref{figure:loss}. 

\begin{table}[t!]
\vspace{2pt}
\centering
\setlength{\tabcolsep}{4pt} 
\renewcommand{\arraystretch}{1.3} 
\begin{center}
\scalebox{1.0}{
    \begin{tabular}{lccc}
        \toprule
        Model & Accuracy & Conciseness & Clarity\\ 
        \hline
        CodeGemma $2$B & $0.516$ & $0.425$ & $91.69$ \\
        CodeGemma $2$B & \multirow{2}{*}{$\bf{0.582}$} & \multirow{2}{*}{$\bf{0.521}$}& \multirow{2}{*}{$\bf{58.75}$} \\
        \vspace{-20pt}\\
        ~Fine-tuned & & &\\
        \bottomrule
    \end{tabular}}
\end{center}
\vspace{5pt}
\caption{Experiment 3 results i.e., comparative evaluations between CodeGemma 2B base model and it's fine-tuned version. The fine-tuned model displays improvements in all metrics compared to the original model.}
\vspace{-20pt}
\label{exp:third}
\end{table}

% Specifically, the fine-tuned CodeGemma achieved an accuracy score of $0.582$, a conciseness score of $0.521$, and a clarity score of $58.75$, compared to the base model's scores of $0.516$, $0.425$, and $91.69$, respectively.

\section{Conclusion and Future Work}
\label{sec:conclusion}

In this project, we introduced DocuMint, a dataset and methodology for evaluating the ability of Small Language Models (SLMs) to generate useful Python docstrings. Our experiments indicate that SLMs are capable of generating docstrings that are accurate, concise, and clear to a reasonable degree. Some of the results are surprising. First, Llama3 8B, a general-purpose SLM, performed the best overall in calculated metrics, despite the fact that its training data only contains a small fraction of code samples compared to code-specific models (although the exact number is not available, it was reported that Llama3 had four times more code samples than the dataset used to train Llama 2, which likely improved its abilities). Second, CodeGemma 7B was perceived as the most accurate by humans, despite having lower accuracy scores in calculated metrics. Further, to the performance of SLMs in generating docstrings, we fine-tuned CodeGemma 2B model on data extracted from open source repositories using World of Code which leads to improvements across all metrics (upto $35\%$ improvement in docstring clarity). 

In the future, there are many interesting directions to extend this work. Expanding DocuMint with more diverse code samples and human-annotated reference docstrings would increase the quality of fine-tuning dataset. Developing better automated metrics to measure docstring quality that better align with human judgement could be another crucial area for future research. To facilitate future efforts, we have release the DocuMint dataset, the fine-tuned CodeGemma $2$B model, and the code we used in this work in public repositories on GitHub and HuggingFace.
\section{Reflections and Lessons Learned}
\label{sec:reflections}

\textbf{Insights into the FOSS Ecosystem:} The process of analyzing the Free and Open Source Software (FOSS) ecosystem through World of Code provided some crucial insights. Initially, we encountered a significant reduction in usable data—from 148,000 extracted functions to only 66,000 actually usable-due to syntactic errors that affected the correct parsing by the AST module. This outlines the challenge that the most popular projects are not always the best written.\\

\noindent \textbf{Subjectivity of Docstring Quality:} Docstring quality in projects varies widely and tends to be subjective, based on what developers deem important to document. For our fine-tuning data, the goal was to find an ideal balance in docstring clarity, conciseness, and accuracy. However, there is no single strategy that helps us achieve this across repositories. \\

\noindent \textbf{Challenges Encountered During Sampling:}
The sampling process itself presented numerous challenges:
\begin{itemize}
\item Syntactic errors in repositories hindered AST parsing.
\item Difficulty in identifying Python-specific files in multi-language repositories.
\item Computational and time expenses associated with parsing and analyzing large codebases.
\item Discrepancies in repository sizes posed challenges in maintaining uniformity in sampling.
\item Subjective differences in the overall quality of the docstrings.
\item Difficulty in avoiding repetitions in such a large dataset (many implementations of the same algorithms might have the same docstrings).
\item Finding a sufficiently diverse set of data.
\end{itemize}

\noindent Further, the human questionnaire survey that we conducted for Experiment $2$ also provided some qualitative insights behind the numerical scores. While the participants had praises for the SLMs output, they also highlighted issues that reduced the quality of the generated docstrings. The most common feedback for all models are listed below:
\vspace{-18pt}
\\
\begin{itemize}
    \item Inconsistencies in spelling, description, and examples.
    \item Incorrect references to arguments and return type.
    \item Raises errors and exceptions not present in code.
    \item Weak explanations for nested loops.
    \item Too long Docstrings.
\end{itemize}
\section{Contributions}
\label{sec:contributions}
All authors contributed equally to the work. \\

\noindent \textbf{Bibek Poudel:} Researched model precision and parameter efficient fine tuning. Introduced HuggingFace and Lightning AI. Secured required compute. Led the fine-tuning effort and technical report writing. Assisted in model inference and evaluation metrics setup.\\

\noindent \textbf{Adam Cook:} Researched available Small Language Models to implement for generating docstrings; developed a Python file that created an output text file containing docstrings generated from an input SLM and JSON data file; calculated the metrics depicted in Table 2.\\

\noindent \textbf{Sekou Traore:} Educated the group on tokenization and data related tasks. Developed a data strategy for fine tuning, familiarized with World of Code and performed data operations such as scraping, filtering, and json formatting as required for fine-tuning.\\

\noindent \textbf{Shelah Ameli:} Initial research on available compute within the ISAAC cluster. Studied standard datasets such as MBPP, HumanEval, and APPS; scraped and filtered a curated dataset for inference. Provided a JSON structure file for setting up inference experiment. Facilitated and compiled the human benchmark experiments.

\begin{acks}
The authors would like to thank Dr. Audris Mockus for his guidance on the project and help with World of Code. The authors would also like to thank the Fluidic City Lab for providing the compute resources. 
\end{acks}

\bibliographystyle{ACM-Reference-Format}
\bibliography{ref}


\begin{thebibliography}{44}

%%% the Web address.

\ifx \showCODEN    \undefined \def \showCODEN     #1{\unskip}     \fi
\ifx \showDOI      \undefined \def \showDOI       #1{#1}\fi
\ifx \showISBNx    \undefined \def \showISBNx     #1{\unskip}     \fi
\ifx \showISBNxiii \undefined \def \showISBNxiii  #1{\unskip}     \fi
\ifx \showISSN     \undefined \def \showISSN      #1{\unskip}     \fi
\ifx \showLCCN     \undefined \def \showLCCN      #1{\unskip}     \fi
\ifx \shownote     \undefined \def \shownote      #1{#1}          \fi
\ifx \showarticletitle \undefined \def \showarticletitle #1{#1}   \fi
\ifx \showURL      \undefined \def \showURL       {\relax}        \fi
% invisible to TeX
\providecommand\bibfield[2]{#2}
\providecommand\bibinfo[2]{#2}
\providecommand\natexlab[1]{#1}
\providecommand\showeprint[2][]{arXiv:#2}

\bibitem[dev(2024)]%
        {devika_github}
 \bibinfo{year}{2024}\natexlab{}.
\newblock \bibinfo{title}{{devika}}.
\newblock \bibinfo{howpublished}{\url{https://github.com/stitionai/devika}}.
\newblock


\bibitem[git(2024)]%
        {github_copilot}
 \bibinfo{year}{2024}\natexlab{}.
\newblock \bibinfo{title}{{GitHub Copilot}}.
\newblock \bibinfo{howpublished}{\url{https://github.com/features/copilot}}.
\newblock


\bibitem[cog(2024)]%
        {cognition_labs}
 \bibinfo{year}{2024}\natexlab{}.
\newblock \bibinfo{title}{{Introducing Devin}}.
\newblock \bibinfo{howpublished}{\url{https://www.cognition-labs.com/introducing-devin}}.
\newblock


\bibitem[ope(2024)]%
        {opendevin_github}
 \bibinfo{year}{2024}\natexlab{}.
\newblock \bibinfo{title}{{OpenDevin}}.
\newblock \bibinfo{howpublished}{\url{https://github.com/OpenDevin/OpenDevin}}.
\newblock


\bibitem[rep(2024)]%
        {replit_ai}
 \bibinfo{year}{2024}\natexlab{}.
\newblock \bibinfo{title}{{Replit AI}}.
\newblock \bibinfo{howpublished}{\url{https://replit.com/ai}}.
\newblock


\bibitem[Abdin et~al\mbox{.}(2024)]%
        {abdin2024phi3}
\bibfield{author}{\bibinfo{person}{Marah Abdin}, \bibinfo{person}{Sam~Ade Jacobs}, {and} \bibinfo{person}{et al.}} \bibinfo{year}{2024}\natexlab{}.
\newblock \showarticletitle{Phi-3 Technical Report: A Highly Capable Language Model Locally on Your Phone}.
\newblock \bibinfo{journal}{\emph{arXiv preprint arXiv:2404.14219}} (\bibinfo{year}{2024}).
\newblock
\urldef\tempurl%
\url{https://arxiv.org/abs/2404.14219}
\showURL{%
\tempurl}
\newblock
\shownote{Accessed 2024 April}.


\bibitem[AI@Meta(2024)]%
        {llama3modelcard}
\bibfield{author}{\bibinfo{person}{AI@Meta}.} \bibinfo{year}{2024}\natexlab{}.
\newblock \showarticletitle{Llama 3 Model Card}.
\newblock  (\bibinfo{year}{2024}).
\newblock
\urldef\tempurl%
\url{https://github.com/meta-llama/llama3/blob/main/MODEL_CARD.md}
\showURL{%
\tempurl}


\bibitem[Anthropic(2024)]%
        {anthropic2024claude3}
\bibfield{author}{\bibinfo{person}{Anthropic}.} \bibinfo{year}{2024}\natexlab{}.
\newblock \bibinfo{title}{Introducing the next generation of Claude}.
\newblock
\newblock
\urldef\tempurl%
\url{https://www.anthropic.com/news/claude-3-family}
\showURL{%
\tempurl}
\newblock
\shownote{Accessed: 2024-05-13}.


\bibitem[Austin et~al\mbox{.}(2021)]%
        {austin2021program}
\bibfield{author}{\bibinfo{person}{Jacob Austin}, \bibinfo{person}{Augustus Odena}, \bibinfo{person}{Maxwell Nye}, \bibinfo{person}{Maarten Bosma}, \bibinfo{person}{Henryk Michalewski}, \bibinfo{person}{David Dohan}, \bibinfo{person}{Ellen Jiang}, \bibinfo{person}{Carrie Cai}, \bibinfo{person}{Michael Terry}, \bibinfo{person}{Quoc Le}, {et~al\mbox{.}}} \bibinfo{year}{2021}\natexlab{}.
\newblock \showarticletitle{Program synthesis with large language models}.
\newblock \bibinfo{journal}{\emph{arXiv preprint arXiv:2108.07732}} (\bibinfo{year}{2021}).
\newblock


\bibitem[Barone and Sennrich(2017)]%
        {barone2017parallel}
\bibfield{author}{\bibinfo{person}{Antonio Valerio~Miceli Barone} {and} \bibinfo{person}{Rico Sennrich}.} \bibinfo{year}{2017}\natexlab{}.
\newblock \showarticletitle{A parallel corpus of python functions and documentation strings for automated code documentation and code generation}.
\newblock \bibinfo{journal}{\emph{arXiv preprint arXiv:1707.02275}} (\bibinfo{year}{2017}).
\newblock


\bibitem[Brusilovsky et~al\mbox{.}(2023)]%
        {brusilovsky2023explaining}
\bibfield{author}{\bibinfo{person}{Peter Brusilovsky}, \bibinfo{person}{Arun-Balajiee Lekshmi-Narayanan}, \bibinfo{person}{Priti Oli}, \bibinfo{person}{Jeevan Chapagain}, \bibinfo{person}{Mohammad Hassany}, \bibinfo{person}{Rabin Banjade}, {and} \bibinfo{person}{Vasile Rus}.} \bibinfo{year}{2023}\natexlab{}.
\newblock \showarticletitle{Explaining code examples in introductory programming courses: Llm vs humans}.
\newblock \bibinfo{journal}{\emph{arXiv preprint arXiv:2403.05538}} (\bibinfo{year}{2023}).
\newblock


\bibitem[Cai et~al\mbox{.}(2024)]%
        {cai2024fly}
\bibfield{author}{\bibinfo{person}{Yufan Cai}, \bibinfo{person}{Yun Lin}, \bibinfo{person}{Chenyan Liu}, \bibinfo{person}{Jinglian Wu}, \bibinfo{person}{Yifan Zhang}, \bibinfo{person}{Yiming Liu}, \bibinfo{person}{Yeyun Gong}, {and} \bibinfo{person}{Jin~Song Dong}.} \bibinfo{year}{2024}\natexlab{}.
\newblock \showarticletitle{On-the-Fly Adapting Code Summarization on Trainable Cost-Effective Language Models}.
\newblock \bibinfo{journal}{\emph{Advances in Neural Information Processing Systems}}  \bibinfo{volume}{36} (\bibinfo{year}{2024}).
\newblock


\bibitem[Chang et~al\mbox{.}(2023)]%
        {chang2023survey}
\bibfield{author}{\bibinfo{person}{Yupeng Chang}, \bibinfo{person}{Xu Wang}, \bibinfo{person}{Jindong Wang}, \bibinfo{person}{Yuan Wu}, \bibinfo{person}{Linyi Yang}, \bibinfo{person}{Kaijie Zhu}, \bibinfo{person}{Hao Chen}, \bibinfo{person}{Xiaoyuan Yi}, \bibinfo{person}{Cunxiang Wang}, \bibinfo{person}{Yidong Wang}, {et~al\mbox{.}}} \bibinfo{year}{2023}\natexlab{}.
\newblock \showarticletitle{A survey on evaluation of large language models}.
\newblock \bibinfo{journal}{\emph{ACM Transactions on Intelligent Systems and Technology}} (\bibinfo{year}{2023}).
\newblock


\bibitem[Chen et~al\mbox{.}(2021)]%
        {chen2021evaluating}
\bibfield{author}{\bibinfo{person}{Mark Chen}, \bibinfo{person}{Jerry Tworek}, \bibinfo{person}{Heewoo Jun}, \bibinfo{person}{Qiming Yuan}, \bibinfo{person}{Henrique Ponde de~Oliveira Pinto}, \bibinfo{person}{Jared Kaplan}, \bibinfo{person}{Harri Edwards}, \bibinfo{person}{Yuri Burda}, \bibinfo{person}{Nicholas Joseph}, \bibinfo{person}{Greg Brockman}, {et~al\mbox{.}}} \bibinfo{year}{2021}\natexlab{}.
\newblock \showarticletitle{Evaluating large language models trained on code}.
\newblock \bibinfo{journal}{\emph{arXiv preprint arXiv:2107.03374}} (\bibinfo{year}{2021}).
\newblock


\bibitem[Devlin et~al\mbox{.}(2018)]%
        {devlin2018bert}
\bibfield{author}{\bibinfo{person}{Jacob Devlin}, \bibinfo{person}{Ming-Wei Chang}, \bibinfo{person}{Kenton Lee}, {and} \bibinfo{person}{Kristina Toutanova}.} \bibinfo{year}{2018}\natexlab{}.
\newblock \showarticletitle{Bert: Pre-training of deep bidirectional transformers for language understanding}.
\newblock \bibinfo{journal}{\emph{arXiv preprint arXiv:1810.04805}} (\bibinfo{year}{2018}).
\newblock


\bibitem[Evtikhiev et~al\mbox{.}(2023)]%
        {evtikhiev2023out}
\bibfield{author}{\bibinfo{person}{Mikhail Evtikhiev}, \bibinfo{person}{Egor Bogomolov}, \bibinfo{person}{Yaroslav Sokolov}, {and} \bibinfo{person}{Timofey Bryksin}.} \bibinfo{year}{2023}\natexlab{}.
\newblock \showarticletitle{Out of the bleu: how should we assess quality of the code generation models?}
\newblock \bibinfo{journal}{\emph{Journal of Systems and Software}}  \bibinfo{volume}{203} (\bibinfo{year}{2023}), \bibinfo{pages}{111741}.
\newblock


\bibitem[{GitHub}(2023)]%
        {github_copilot_business}
\bibfield{author}{\bibinfo{person}{{GitHub}}.} \bibinfo{year}{2023}\natexlab{}.
\newblock \bibinfo{title}{{GitHub Copilot for Business is now available}}.
\newblock \bibinfo{howpublished}{\url{https://github.blog/2023-02-14-github-copilot-for-business-is-now-available/}}.
\newblock


\bibitem[{Google}(2024)]%
        {code_gemma}
\bibfield{author}{\bibinfo{person}{{Google}}.} \bibinfo{year}{2024}\natexlab{}.
\newblock \bibinfo{title}{CodeGemma: Open Code Models Based on Gemma}.
\newblock \bibinfo{howpublished}{\url{https://storage.googleapis.com/deepmind-media/gemma/codegemma_report.pdf}}.
\newblock


\bibitem[Google(2024)]%
        {google2024gemma}
\bibfield{author}{\bibinfo{person}{Google}.} \bibinfo{year}{2024}\natexlab{}.
\newblock \bibinfo{title}{Gemma: Open Models for Developers}.
\newblock \bibinfo{howpublished}{\url{https://blog.google/technology/developers/gemma-open-models/}}.
\newblock


\bibitem[Hendrycks et~al\mbox{.}(2021)]%
        {hendrycks2021measuring}
\bibfield{author}{\bibinfo{person}{Dan Hendrycks}, \bibinfo{person}{Steven Basart}, \bibinfo{person}{Saurav Kadavath}, \bibinfo{person}{Mantas Mazeika}, \bibinfo{person}{Akul Arora}, \bibinfo{person}{Ethan Guo}, \bibinfo{person}{Collin Burns}, \bibinfo{person}{Samir Puranik}, \bibinfo{person}{Horace He}, \bibinfo{person}{Dawn Song}, {et~al\mbox{.}}} \bibinfo{year}{2021}\natexlab{}.
\newblock \showarticletitle{Measuring coding challenge competence with apps}.
\newblock \bibinfo{journal}{\emph{arXiv preprint arXiv:2105.09938}} (\bibinfo{year}{2021}).
\newblock


\bibitem[Hu et~al\mbox{.}(2021)]%
        {hu2021lora}
\bibfield{author}{\bibinfo{person}{Edward~J Hu}, \bibinfo{person}{Yelong Shen}, \bibinfo{person}{Phillip Wallis}, \bibinfo{person}{Zeyuan Allen-Zhu}, \bibinfo{person}{Yuanzhi Li}, \bibinfo{person}{Shean Wang}, \bibinfo{person}{Lu Wang}, {and} \bibinfo{person}{Weizhu Chen}.} \bibinfo{year}{2021}\natexlab{}.
\newblock \showarticletitle{Lora: Low-rank adaptation of large language models}.
\newblock \bibinfo{journal}{\emph{arXiv preprint arXiv:2106.09685}} (\bibinfo{year}{2021}).
\newblock


\bibitem[Hu et~al\mbox{.}(2018)]%
        {hu2018deep}
\bibfield{author}{\bibinfo{person}{Xing Hu}, \bibinfo{person}{Ge Li}, \bibinfo{person}{Xin Xia}, \bibinfo{person}{David Lo}, {and} \bibinfo{person}{Zhi Jin}.} \bibinfo{year}{2018}\natexlab{}.
\newblock \showarticletitle{Deep code comment generation}. In \bibinfo{booktitle}{\emph{Proceedings of the 26th conference on program comprehension}}. \bibinfo{pages}{200--210}.
\newblock


\bibitem[Jimenez et~al\mbox{.}(2023)]%
        {jimenez2023swe}
\bibfield{author}{\bibinfo{person}{Carlos~E Jimenez}, \bibinfo{person}{John Yang}, \bibinfo{person}{Alexander Wettig}, \bibinfo{person}{Shunyu Yao}, \bibinfo{person}{Kexin Pei}, \bibinfo{person}{Ofir Press}, {and} \bibinfo{person}{Karthik Narasimhan}.} \bibinfo{year}{2023}\natexlab{}.
\newblock \showarticletitle{Swe-bench: Can language models resolve real-world github issues?}
\newblock \bibinfo{journal}{\emph{arXiv preprint arXiv:2310.06770}} (\bibinfo{year}{2023}).
\newblock


\bibitem[Kaddour et~al\mbox{.}(2023)]%
        {kaddour2023challenges}
\bibfield{author}{\bibinfo{person}{Jean Kaddour}, \bibinfo{person}{Joshua Harris}, \bibinfo{person}{Maximilian Mozes}, \bibinfo{person}{Herbie Bradley}, \bibinfo{person}{Roberta Raileanu}, {and} \bibinfo{person}{Robert McHardy}.} \bibinfo{year}{2023}\natexlab{}.
\newblock \showarticletitle{Challenges and applications of large language models}.
\newblock \bibinfo{journal}{\emph{arXiv preprint arXiv:2307.10169}} (\bibinfo{year}{2023}).
\newblock


\bibitem[Kojima et~al\mbox{.}(2022)]%
        {kojima2022large}
\bibfield{author}{\bibinfo{person}{Takeshi Kojima}, \bibinfo{person}{Shixiang~Shane Gu}, \bibinfo{person}{Machel Reid}, \bibinfo{person}{Yutaka Matsuo}, {and} \bibinfo{person}{Yusuke Iwasawa}.} \bibinfo{year}{2022}\natexlab{}.
\newblock \showarticletitle{Large language models are zero-shot reasoners}.
\newblock \bibinfo{journal}{\emph{Advances in neural information processing systems}}  \bibinfo{volume}{35} (\bibinfo{year}{2022}), \bibinfo{pages}{22199--22213}.
\newblock


\bibitem[Li et~al\mbox{.}(2022)]%
        {li2022competition}
\bibfield{author}{\bibinfo{person}{Yujia Li}, \bibinfo{person}{David Choi}, \bibinfo{person}{Junyoung Chung}, \bibinfo{person}{Nate Kushman}, \bibinfo{person}{Julian Schrittwieser}, \bibinfo{person}{R{\'e}mi Leblond}, \bibinfo{person}{Tom Eccles}, \bibinfo{person}{James Keeling}, \bibinfo{person}{Felix Gimeno}, \bibinfo{person}{Agustin Dal~Lago}, {et~al\mbox{.}}} \bibinfo{year}{2022}\natexlab{}.
\newblock \showarticletitle{Competition-level code generation with alphacode}.
\newblock \bibinfo{journal}{\emph{Science}} \bibinfo{volume}{378}, \bibinfo{number}{6624} (\bibinfo{year}{2022}), \bibinfo{pages}{1092--1097}.
\newblock


\bibitem[Liu et~al\mbox{.}(2023)]%
        {liu2023your}
\bibfield{author}{\bibinfo{person}{Jiawei Liu}, \bibinfo{person}{Chunqiu~Steven Xia}, \bibinfo{person}{Yuyao Wang}, {and} \bibinfo{person}{Lingming Zhang}.} \bibinfo{year}{2023}\natexlab{}.
\newblock \showarticletitle{Is your code generated by ChatGPT really correct}.
\newblock \bibinfo{journal}{\emph{Rigorous evaluation of large language models for code generation. CoRR, abs/2305.01210}} (\bibinfo{year}{2023}).
\newblock


\bibitem[Lozhkov et~al\mbox{.}(2024)]%
        {lozhkov2024starcoder}
\bibfield{author}{\bibinfo{person}{Anton Lozhkov}, \bibinfo{person}{Raymond Li}, \bibinfo{person}{Loubna~Ben Allal}, \bibinfo{person}{Federico Cassano}, \bibinfo{person}{Joel Lamy-Poirier}, \bibinfo{person}{Nouamane Tazi}, \bibinfo{person}{Ao Tang}, \bibinfo{person}{Dmytro Pykhtar}, \bibinfo{person}{Jiawei Liu}, \bibinfo{person}{Yuxiang Wei}, {et~al\mbox{.}}} \bibinfo{year}{2024}\natexlab{}.
\newblock \showarticletitle{StarCoder 2 and The Stack v2: The Next Generation}.
\newblock \bibinfo{journal}{\emph{arXiv preprint arXiv:2402.19173}} (\bibinfo{year}{2024}).
\newblock


\bibitem[Luo et~al\mbox{.}(2024)]%
        {luo2024repoagent}
\bibfield{author}{\bibinfo{person}{Qinyu Luo}, \bibinfo{person}{Yining Ye}, \bibinfo{person}{Shihao Liang}, \bibinfo{person}{Zhong Zhang}, \bibinfo{person}{Yujia Qin}, \bibinfo{person}{Yaxi Lu}, \bibinfo{person}{Yesai Wu}, \bibinfo{person}{Xin Cong}, \bibinfo{person}{Yankai Lin}, \bibinfo{person}{Yingli Zhang}, {et~al\mbox{.}}} \bibinfo{year}{2024}\natexlab{}.
\newblock \showarticletitle{RepoAgent: An LLM-Powered Open-Source Framework for Repository-level Code Documentation Generation}.
\newblock \bibinfo{journal}{\emph{arXiv preprint arXiv:2402.16667}} (\bibinfo{year}{2024}).
\newblock


\bibitem[Ma et~al\mbox{.}(2019)]%
        {ma2019world}
\bibfield{author}{\bibinfo{person}{Yuxing Ma}, \bibinfo{person}{Chris Bogart}, \bibinfo{person}{Sadika Amreen}, \bibinfo{person}{Russell Zaretzki}, {and} \bibinfo{person}{Audris Mockus}.} \bibinfo{year}{2019}\natexlab{}.
\newblock \showarticletitle{World of code: an infrastructure for mining the universe of open source VCS data}. In \bibinfo{booktitle}{\emph{2019 IEEE/ACM 16th International Conference on Mining Software Repositories (MSR)}}. IEEE, \bibinfo{pages}{143--154}.
\newblock


\bibitem[Ma et~al\mbox{.}(2021)]%
        {ma2021world}
\bibfield{author}{\bibinfo{person}{Yuxing Ma}, \bibinfo{person}{Tapajit Dey}, \bibinfo{person}{Chris Bogart}, \bibinfo{person}{Sadika Amreen}, \bibinfo{person}{Marat Valiev}, \bibinfo{person}{Adam Tutko}, \bibinfo{person}{David Kennard}, \bibinfo{person}{Russell Zaretzki}, {and} \bibinfo{person}{Audris Mockus}.} \bibinfo{year}{2021}\natexlab{}.
\newblock \showarticletitle{World of code: enabling a research workflow for mining and analyzing the universe of open source VCS data}.
\newblock \bibinfo{journal}{\emph{Empirical Software Engineering}}  \bibinfo{volume}{26} (\bibinfo{year}{2021}), \bibinfo{pages}{1--42}.
\newblock


\bibitem[McBurney and McMillan(2014)]%
        {mcburney2014automatic}
\bibfield{author}{\bibinfo{person}{Paul~W McBurney} {and} \bibinfo{person}{Collin McMillan}.} \bibinfo{year}{2014}\natexlab{}.
\newblock \showarticletitle{Automatic documentation generation via source code summarization of method context}. In \bibinfo{booktitle}{\emph{Proceedings of the 22nd International Conference on Program Comprehension}}. \bibinfo{pages}{279--290}.
\newblock


\bibitem[{Meta}(2023)]%
        {meta_llama_2023}
\bibfield{author}{\bibinfo{person}{{Meta}}.} \bibinfo{year}{2023}\natexlab{}.
\newblock \bibinfo{title}{LLaMA}.
\newblock \bibinfo{howpublished}{\url{https://llama.meta.com}}.
\newblock


\bibitem[{Microsoft Research}(2024)]%
        {microsoft_phi_2}
\bibfield{author}{\bibinfo{person}{{Microsoft Research}}.} \bibinfo{year}{2024}\natexlab{}.
\newblock \bibinfo{title}{Phi-2: The Surprising Power of Small Language Models}.
\newblock \bibinfo{howpublished}{\url{https://www.microsoft.com/en-us/research/blog/\\phi-2-the-surprising-power-of-\\small-language-models/}}.
\newblock


\bibitem[Nijkamp et~al\mbox{.}(2022)]%
        {nijkamp2022codegen}
\bibfield{author}{\bibinfo{person}{Erik Nijkamp}, \bibinfo{person}{Bo Pang}, \bibinfo{person}{Hiroaki Hayashi}, \bibinfo{person}{Lifu Tu}, \bibinfo{person}{Huan Wang}, \bibinfo{person}{Yingbo Zhou}, \bibinfo{person}{Silvio Savarese}, {and} \bibinfo{person}{Caiming Xiong}.} \bibinfo{year}{2022}\natexlab{}.
\newblock \showarticletitle{Codegen: An open large language model for code with multi-turn program synthesis}.
\newblock \bibinfo{journal}{\emph{arXiv preprint arXiv:2203.13474}} (\bibinfo{year}{2022}).
\newblock


\bibitem[Roziere et~al\mbox{.}(2023)]%
        {roziere2023code}
\bibfield{author}{\bibinfo{person}{Baptiste Roziere}, \bibinfo{person}{Jonas Gehring}, \bibinfo{person}{Fabian Gloeckle}, \bibinfo{person}{Sten Sootla}, \bibinfo{person}{Itai Gat}, \bibinfo{person}{Xiaoqing~Ellen Tan}, \bibinfo{person}{Yossi Adi}, \bibinfo{person}{Jingyu Liu}, \bibinfo{person}{Tal Remez}, \bibinfo{person}{J{\'e}r{\'e}my Rapin}, {et~al\mbox{.}}} \bibinfo{year}{2023}\natexlab{}.
\newblock \showarticletitle{Code llama: Open foundation models for code}.
\newblock \bibinfo{journal}{\emph{arXiv preprint arXiv:2308.12950}} (\bibinfo{year}{2023}).
\newblock


\bibitem[Sarsa et~al\mbox{.}(2022)]%
        {sarsa2022automatic}
\bibfield{author}{\bibinfo{person}{Sami Sarsa}, \bibinfo{person}{Paul Denny}, \bibinfo{person}{Arto Hellas}, {and} \bibinfo{person}{Juho Leinonen}.} \bibinfo{year}{2022}\natexlab{}.
\newblock \showarticletitle{Automatic generation of programming exercises and code explanations using large language models}. In \bibinfo{booktitle}{\emph{Proceedings of the 2022 ACM Conference on International Computing Education Research-Volume 1}}. \bibinfo{pages}{27--43}.
\newblock


\bibitem[Taori et~al\mbox{.}(2023)]%
        {taori2023alpaca}
\bibfield{author}{\bibinfo{person}{Rohan Taori}, \bibinfo{person}{Ishaan Gulrajani}, \bibinfo{person}{Tianyi Zhang}, \bibinfo{person}{Yann Dubois}, \bibinfo{person}{Xuechen Li}, \bibinfo{person}{Carlos Guestrin}, \bibinfo{person}{Percy Liang}, {and} \bibinfo{person}{Tatsunori~B Hashimoto}.} \bibinfo{year}{2023}\natexlab{}.
\newblock \showarticletitle{Alpaca: A strong, replicable instruction-following model}.
\newblock \bibinfo{journal}{\emph{Stanford Center for Research on Foundation Models. https://crfm. stanford. edu/2023/03/13/alpaca. html}} \bibinfo{volume}{3}, \bibinfo{number}{6} (\bibinfo{year}{2023}), \bibinfo{pages}{7}.
\newblock


\bibitem[Wang et~al\mbox{.}(2024)]%
        {wang2024demo2code}
\bibfield{author}{\bibinfo{person}{Yuki Wang}, \bibinfo{person}{Gonzalo Gonzalez-Pumariega}, \bibinfo{person}{Yash Sharma}, {and} \bibinfo{person}{Sanjiban Choudhury}.} \bibinfo{year}{2024}\natexlab{}.
\newblock \showarticletitle{Demo2code: From summarizing demonstrations to synthesizing code via extended chain-of-thought}.
\newblock \bibinfo{journal}{\emph{Advances in Neural Information Processing Systems}}  \bibinfo{volume}{36} (\bibinfo{year}{2024}).
\newblock


\bibitem[Wei et~al\mbox{.}(2022)]%
        {wei2022chain}
\bibfield{author}{\bibinfo{person}{Jason Wei}, \bibinfo{person}{Xuezhi Wang}, \bibinfo{person}{Dale Schuurmans}, \bibinfo{person}{Maarten Bosma}, \bibinfo{person}{Fei Xia}, \bibinfo{person}{Ed Chi}, \bibinfo{person}{Quoc~V Le}, \bibinfo{person}{Denny Zhou}, {et~al\mbox{.}}} \bibinfo{year}{2022}\natexlab{}.
\newblock \showarticletitle{Chain-of-thought prompting elicits reasoning in large language models}.
\newblock \bibinfo{journal}{\emph{Advances in neural information processing systems}}  \bibinfo{volume}{35} (\bibinfo{year}{2022}), \bibinfo{pages}{24824--24837}.
\newblock


\bibitem[Xia et~al\mbox{.}(2017)]%
        {xia2017measuring}
\bibfield{author}{\bibinfo{person}{Xin Xia}, \bibinfo{person}{Lingfeng Bao}, \bibinfo{person}{David Lo}, \bibinfo{person}{Zhenchang Xing}, \bibinfo{person}{Ahmed~E Hassan}, {and} \bibinfo{person}{Shanping Li}.} \bibinfo{year}{2017}\natexlab{}.
\newblock \showarticletitle{Measuring program comprehension: A large-scale field study with professionals}.
\newblock \bibinfo{journal}{\emph{IEEE Transactions on Software Engineering}} \bibinfo{volume}{44}, \bibinfo{number}{10} (\bibinfo{year}{2017}), \bibinfo{pages}{951--976}.
\newblock


\bibitem[Zhang et~al\mbox{.}(2019)]%
        {zhang2019bertscore}
\bibfield{author}{\bibinfo{person}{Tianyi Zhang}, \bibinfo{person}{Varsha Kishore}, \bibinfo{person}{Felix Wu}, \bibinfo{person}{Kilian~Q Weinberger}, {and} \bibinfo{person}{Yoav Artzi}.} \bibinfo{year}{2019}\natexlab{}.
\newblock \showarticletitle{Bertscore: Evaluating text generation with bert}.
\newblock \bibinfo{journal}{\emph{arXiv preprint arXiv:1904.09675}} (\bibinfo{year}{2019}).
\newblock


\bibitem[Zhang et~al\mbox{.}(2023)]%
        {zhang2023unifying}
\bibfield{author}{\bibinfo{person}{Ziyin Zhang}, \bibinfo{person}{Chaoyu Chen}, \bibinfo{person}{Bingchang Liu}, \bibinfo{person}{Cong Liao}, \bibinfo{person}{Zi Gong}, \bibinfo{person}{Hang Yu}, \bibinfo{person}{Jianguo Li}, {and} \bibinfo{person}{Rui Wang}.} \bibinfo{year}{2023}\natexlab{}.
\newblock \showarticletitle{Unifying the perspectives of nlp and software engineering: A survey on language models for code}.
\newblock \bibinfo{journal}{\emph{arXiv preprint arXiv:2311.07989}} (\bibinfo{year}{2023}).
\newblock


\bibitem[Zhao et~al\mbox{.}(2023)]%
        {zhao2023survey}
\bibfield{author}{\bibinfo{person}{Wayne~Xin Zhao}, \bibinfo{person}{Kun Zhou}, \bibinfo{person}{Junyi Li}, \bibinfo{person}{Tianyi Tang}, \bibinfo{person}{Xiaolei Wang}, \bibinfo{person}{Yupeng Hou}, \bibinfo{person}{Yingqian Min}, \bibinfo{person}{Beichen Zhang}, \bibinfo{person}{Junjie Zhang}, \bibinfo{person}{Zican Dong}, {et~al\mbox{.}}} \bibinfo{year}{2023}\natexlab{}.
\newblock \showarticletitle{A survey of large language models}.
\newblock \bibinfo{journal}{\emph{arXiv preprint arXiv:2303.18223}} (\bibinfo{year}{2023}).
\newblock


\end{thebibliography}
\clearpage
% \newpage
\appendix
\section{Appendix}
\subsection{Experiment 2 Human Questionnaire Sample}
\label{appendix}


\section*{Supplementary material}

# A Appendix
## A.1 Experiment 2 Human Questionnaire Sample

**Docu-Mint: Evaluation of Docstring Generation for Python using Small Language Models**

<u>**Instructions**</u>:

Your goal is to analyze the following **docstrings** (string literals for code documentation) and evaluate them based on personal bias. You will use the following metrics:
- **Accuracy**: the docstring contains all necessary inputs, outputs, etc.
- **Concise**: the docstring conveys information succinctly
- **Clear**: the docstring is easy to read and understand

Below is an example function and docstring that meets the above metrics:

```python
def odd_values_string(str):
    result = \"\"
    for i in range(len(str)):
        if i % 2 == 0:
            result = result + str[i]
    return result
"""
    Extracts characters at odd indices from a string.

    Parameters:
    str (str): The input string.

    Returns:
    str: A new string containing only the characters at odd indices of the input string.

    Example:
    odd_values_string("abcdef") returns "ace"
    odd_values_string("python") returns "pto"
"""
```

<u>**Please fill out the following information below:**</u>

Name: _____

Age: ____

Do you have experience writing code in Python? Yes / No

If so, how many years of experience do you have? _____

How often do you read docstrings?

1. Daily
2. Several times a week
3. Once a week
4. Several times a month
5. Once a month
6. Never

**Question 1:**

```
1  def largest_neg(list1):
2      max = list1[0]
3      for x in list1:
4          if x < max:
5              max = x
6      return max
```
"""
Finds the largest negative number in a list.

Args:
    list1: A list of numbers.

Returns:
    The largest negative number in the list.

Raises:
    ValueError: If the list is empty.
"""

On a scale of 1 (poor) to 10 (excellent):

Rate the docstring based on how **accurate** it is:

1   2   3   4   5   6   7   8   9   10

Rate the docstring based on how **concise** it is:

1   2   3   4   5   6   7   8   9   10

Rate the docstring based on how **clear** it is:

1   2   3   4   5   6   7   8   9   10

How would you rate the overall quality of the docstring?

1   2   3   4   5   6   7   8   9   10

Please provide any additional comments



\end{document}